\documentclass[useAMS,usenatbib]{mn2e}

\usepackage[dvips]{graphicx}  
\usepackage{color}
\usepackage{aas_macros}  
\usepackage{amsmath}
\usepackage{amssymb}

\newcommand{\Msolar}{\mbox{\,$\rm M_{\odot}$}}        
\newcommand{\Rsolar}{\mbox{\,$\rm R_{\odot}$}}        

  \newcommand{\Teff}{\mbox{\,\em T$_{\rm eff}$}}         

  \newcommand{\ion}[2]{\mbox{\,#1\,{\sc #2}}}         
  \newcommand{\kmsec}{\,\mbox{$\mbox{km}\,\mbox{s}^{-1}$}}    
  \newcommand{\cmss}{\,\mbox{$\mbox{cm}\,\mbox{s}^{-2}$}}    

  \def\simge{\mathrel{\raise1.16pt\hbox{$>$}\kern-7.0pt
    \lower3.06pt\hbox{{$\scriptstyle \sim$}}}}           
  \def\simle{\mathrel{\raise1.16pt\hbox{$<$}\kern-7.0pt
    \lower3.06pt\hbox{{$\scriptstyle \sim$}}}}           
   \newcommand{\lsfour}{\mbox{LS\,IV\,$-14^{\circ}116$}}                  

\title[LS IV$-14^{\circ}116$: radial velocities]{Radial velocity measurements of the pulsating zirconium star: 
LS IV$-14^{\circ}116$\footnote{Based on data collected with the William Herschel Telescope,
 the Anglo-Australian Telescope, the SAAO 1.9\,m telescope, the ANU 2.3\,m 
telescope,  and the ESO Very Large Telescope. } }
\author[C. S. Jeffery et al.]{C. S. Jeffery$^{1,2}$\footnote{email: csj@arm.ac.uk}, A. Ahmad$^1$, Naslim, N.$^{1,3}$, W. Kerzendorf$^{4,5}$ \\
$^{1}$Armagh Observatory, College Hill, Armagh BT61 9DG, UK\\
$^{2}$School of Physics, Trinity College Dublin, College Green, Dublin 2, Ireland\\
$^{3}$Academia Sinica Institute of Astronomy and Astrophysics,  Taipei 10617, Taiwan R.O.C\\
$^{4}$Research School of Astronomy \& Astrophysics,
ANU College of Physical \& Mathematical Sciences,
Mount Stromlo Observatory, \\ACT 2611, 
Australia \\
$^{5}$Dept of Astronomy and Astrophysics,
University of Toronto,
50 St. George Street, 
Toronto, Ontario, Canada, M5S 3H4
}

\begin{document}

\date{Accepted \ldots. Received \ldots; in original form \ldots}

\pagerange{\pageref{firstpage}--\pageref{lastpage}} \pubyear{2014}

\maketitle

\label{firstpage}

\begin{abstract}
The helium-rich hot subdwarf \lsfour\ shows remarkably high surface abundances of 
zirconium, yttrium, strontium, and germanium, indicative of strong chemical stratification in 
the photosphere. It also shows photometric behaviour indicative of non-radial g-mode
pulsations, despite having surface properties inconsistent with any known 
pulsational instability zone.  We have conducted a search for radial velocity variability.
This has demonstrated that at least one photometric period is observable in several 
absorption lines as a radial velocity variation with a semi-amplitude in excess of 5\kmsec.
A correlation between line strength and pulsation amplitude provides evidence 
that the photosphere pulsates differentially. The ratio of light to velocity 
amplitude is too small to permit the largest amplitude oscillation to be radial. 
\end{abstract}


\section{Introduction}
\label{s:intro}

The helium-rich hot subdwarf \lsfour\ is remarkable for several reasons. Its  peculiarity was 
first recognised  in a spectroscopic follow-up of UV-excess  objects \citep{viton91}. Subsequent analyses
confirmed the excess helium abundance and high effective temperature (\Teff), making it one 
of  relatively few
such stars \citep{ahmad03,ahmad04}, and the only one to show photometric variability \citep{ahmad05}. 
Follow-up photometry confirmed the star to be a multi-periodic variable, almost
certainly a non-radial g-mode oscillator \citep{jeffery11.ibvs,green11} with a dominant pulsation 
signal at 1953\,s (=0.022\,d) and weaker signals at 2620, 2872, 3582, 4260 and 5084\,s. However 
\lsfour\ presents a problem in that 
its effective temperature and gravity  are deemed inconsistent with domains
known to be unstable to g-mode oscillations \citep{jeffery06a,jeffery07,green11}. 
A further surprise came with  the discovery of superabundances (by $\sim 4$ dex) 
of zirconium, yttrium, strontium, 
and, to a lesser extent, germanium \citep{naslim11}. It has been argued that these could be
associated with a dynamical self-stratification of the photosphere as a young subdwarf 
contracts towards the extreme horizontal branch  \citep{naslim13},  suggesting that 
\lsfour\ is a proto-sdB star which will eventually develop a helium-poor atmosphere,  
and prompting  speculation that the oscillations might represent the first known case of 
$\epsilon-$driven\footnote{{\it i.e.} driven by instability in a nuclear burning zone.} 
pulsation \citep{bertolami11,bertolami12.liege}.  Meanwhile, the exotic surface
chemistry suggested that, as in some Bp(He) stars \citep{townsend05}, 
magnetic activity might be responsible for the variability \citep{naslim11},  a speculation quite 
delicately  dismantled by \citet{green11}. 

Whilst \lsfour\  presents conundra in terms
of the driving mechanism for the pulsation, its surface composition and its actual effective temperature \citep{green11}, 
more basic data are also required. First, many hot subdwarfs are known to be members 
of binaries, providing at least three mechanisms for the removal of  surface layers 
necessary for evolution to the extreme horizontal branch. \lsfour\ is a very slow rotator \citep{naslim11};
would radial-velocity measurements show it to be a binary?  
Second, if light variations are due to pulsation (and not magnetic activity), surface motion should
be detectable in radial-velocity data at periods equal to the light variations. Such
measurements could ultimately provide direct radius measurements or  mode identifications 
using techniques similar to  the Baade-Wesselink method \citep{stamford81}. Third,
\citet{naslim11} suggested that, if  the photosphere of \lsfour\ is chemically stratified and pulsating, 
 differential motion could be detected by comparing the radial velocity curves 
due to different elements.
As a consequence, we have  attempted over several years to resolve the surface motion of 
\lsfour. This paper describes the observations attempted and the ultimate detection of 
surface motion  associated with the principal photometric oscillation at a period of $\sim1950$\,s.

\begin{center}
\begin{table*}
\caption{Observing log.}
\label{t:log}
\begin{tabular}{l lc cc rrlll}
\hline
 Telescope / & Date &  UT$_{\rm start}$ --  UT$_{\rm end}$ &  $\lambda$  & $R$ & $t_{\rm exp}$ & $n_{\rm exp}$ & $\langle$S/N$\rangle$ & $\langle v_{\rm T} \rangle $ & Observer\\
Instrument &            & & nm &    & s &     &     & \kmsec & \\
\hline
WHT/ISIS             &  2005 May 27        &  03:30 -- 05:00  & 436--507  &  10\,000  &  300 & 8   &  100  &  $-153\pm1$  & Ahmad \\
WHT/ISIS             &  2005 May 28        &  04:20 -- 05:00 &  376--452  &   9\,000  &    45 & 32 & 15--40 &  $-153\pm1$  & Ahmad \\
AAT/UCLES$^1$  &  2005 Aug 27         & 11:41 -- 13:30 &   382--521  & 30\,000 & 1800 & 3  & 40 & $-153\pm1$ & Ahmad \\
SAAO\,1.9/CGS  &  2007 June 29         & 01:03 -- 04:15 &  380--560  &  2\,000  &  300 &  36 & 20--30 &  $(-184\pm10)$  & Ahmad  \\
SAAO\,1.9/CGS  &  2007 June 29--30 & 23:43 -- 02:54 &  "   &  2\,000  &  " &  36 & " &   " & Ahmad  \\
SAAO\,1.9/CGS  &  2007 July 02--03  & 23:39 -- 03:56 &  "  &  2\,000  &  " &  48  & " &  "  & Ahmad \\
SAAO\,1.9/CGS  &  2007 July 03--04  & 23:46 -- 04:03 &  "  &  2\,000  &  " &  48 &  " &  "  & Ahmad  \\
ANU\,2.3/WiFES  &   2010  June 18       & 12:42 -- 14:33 & 418--552  &  5\,000   &    480    & 9   & 100 & $(-137\pm3)$  & Kerzendorf \\
VLT/UVES            &  2011 Sept 07        & 02:31 -- 06:20 & 330--680   & 100\,000  &  300 & 39 & 30--40 & $-155\pm1$  & Service \\ 
\hline
\multicolumn{5}{l}{1: \citet{naslim11}} & & & & \\
\end{tabular}
\end{table*}
\end{center}

\section[]{Observations}

Spectroscopic time-series observations of \lsfour\ were obtained in 2005 with the 
Intermediate dispersion Spectrograph and Imaging System (ISIS)  on the 
William Herschel Telescope (WHT), in 2005 with the University College Echelle Spectrograph (UCLES) of the 
Anglo-Australian Telescope (AAT), in 2007 with the {\bf  Cassegrain Grating Spectrograph (CGS) } on 
the 1.9\,m telescope of the South African Astronomical Observatory (SAAO\,1.9),
in 2010 with the Wide-Field Spectrograph (WiFeS) on the Australian National University 2.3\,m telescope (ANU\,2.3), 
and in 2011 with the Ultraviolet and Visual Echelle Spectrograph (UVES) 
on the Very Large Telescope (VLT) of the European Southern Observatory (ESO).
Details of dates observed, duration of observations in UT, 
wavelength range covered $\lambda\lambda$, spectral resolution $R$, 
exposure times  $t_{\rm exp}$, numbers of exposures $n_{\rm exp}$, and average signal-to-noise ratio
per exposure $\langle$S/N$\rangle$ are given in Table\,\ref{t:log}.
The spectra were reduced using standard 
procedures, including  bias-subtraction, flat-fielding, sky-subtraction, wavelength-calibration, and 
rectification. For UVES, the  ESO pipeline reduced data were recovered from the ESO archive. 

\begin{figure}
	\centering
		\includegraphics[width=0.47\textwidth]{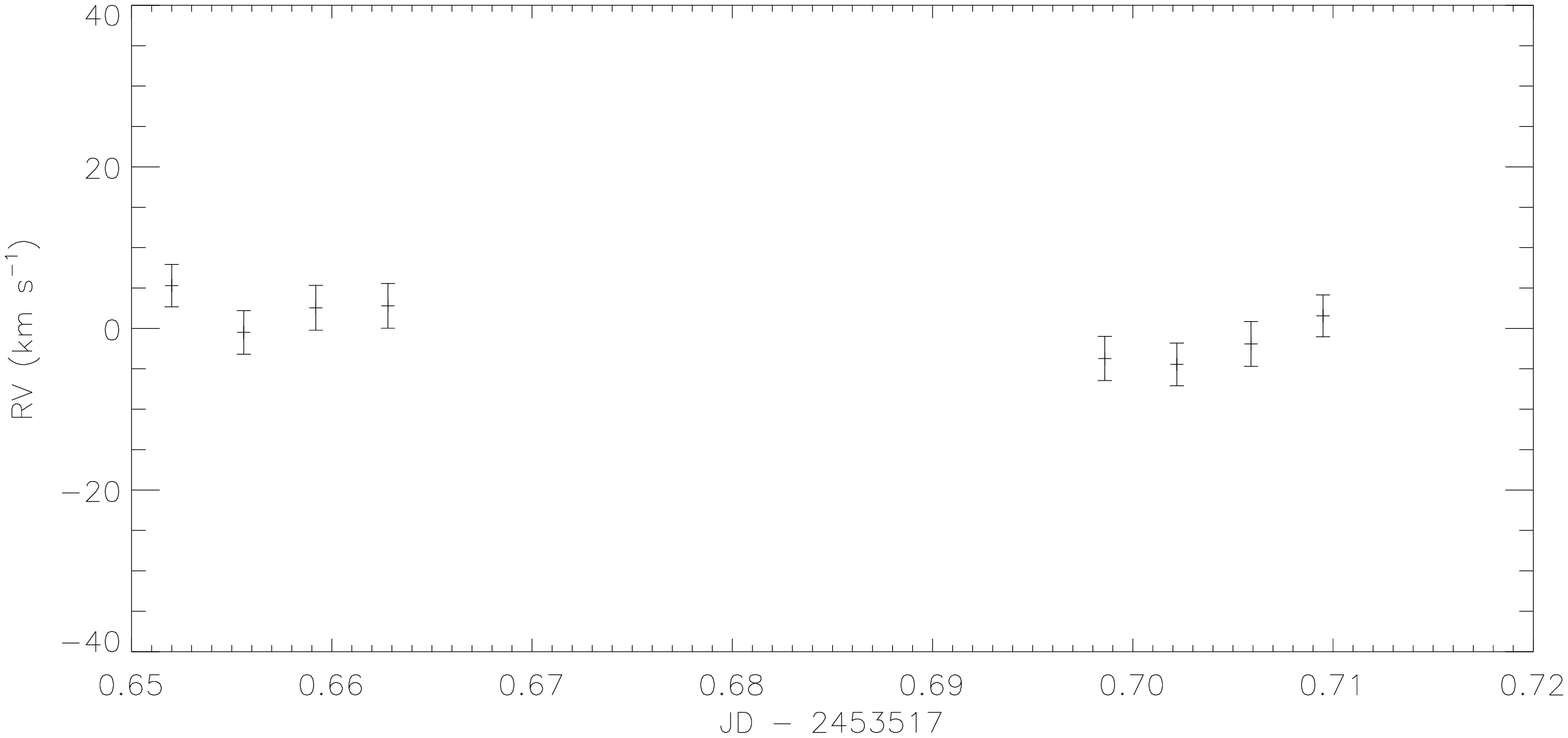}\\[1mm]
		\includegraphics[width=0.47\textwidth]{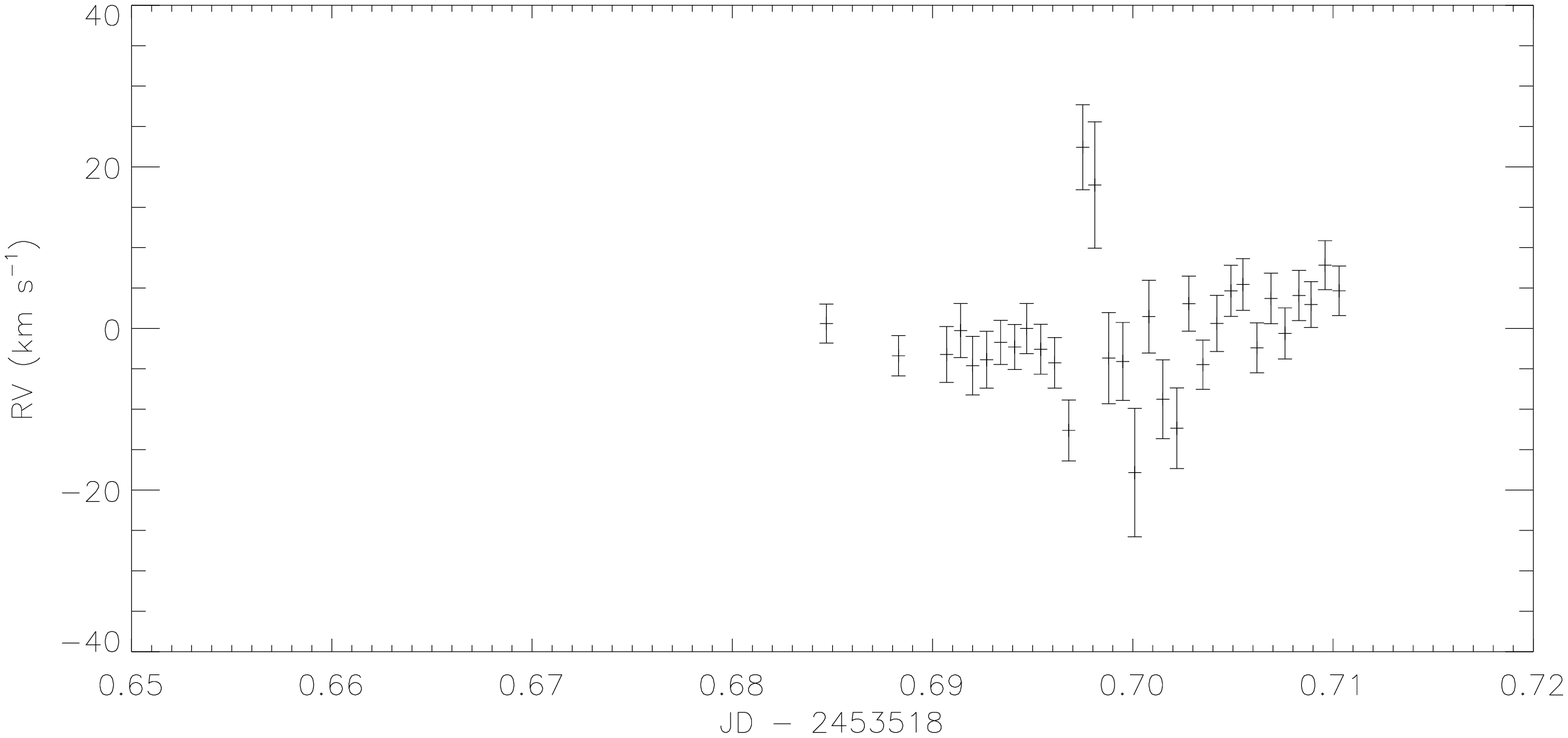}
\caption{Relative radial velocities for \lsfour\ obtained with  WHT/ISIS on 2005 May 27 and 28.} 
	\label{f:wht}
\end{figure}

\begin{figure}
	\centering
		\includegraphics[width=0.47\textwidth]{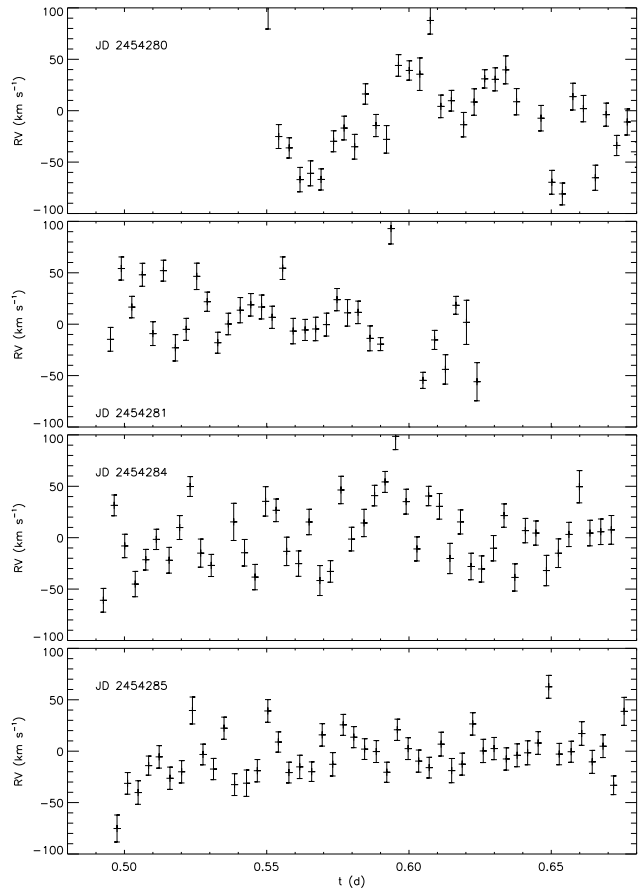}
\caption{Relative radial velocities for \lsfour\ obtained with SAAO\,1.9/CGS on 2007 June 29 - July 04.} 
	\label{f:saao}
\end{figure}

\begin{figure}
	\centering
		\includegraphics[width=0.47\textwidth]{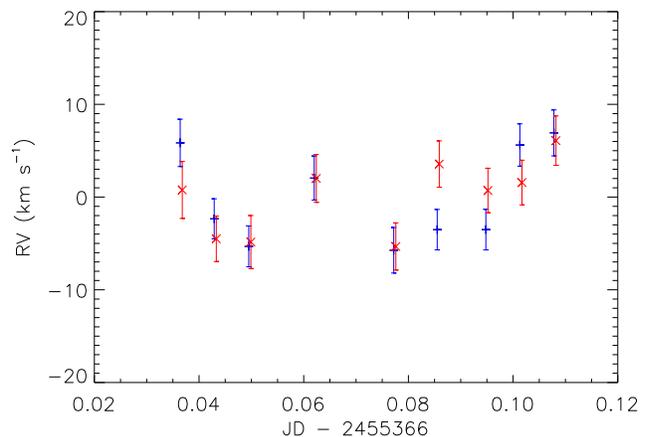}
\caption{Relative radial velocities for \lsfour\ obtained with ANU\,2.3/WiFES on 2010 June 18 (blue `$+$' 420--488\,nm, red `$\times$' 490--550\,nm -- offset by $\pm0.0002$\, d for clarity).} 
	\label{f:anu}
\end{figure}

\begin{figure}
	\centering
		\includegraphics[width=0.47\textwidth,angle=0]{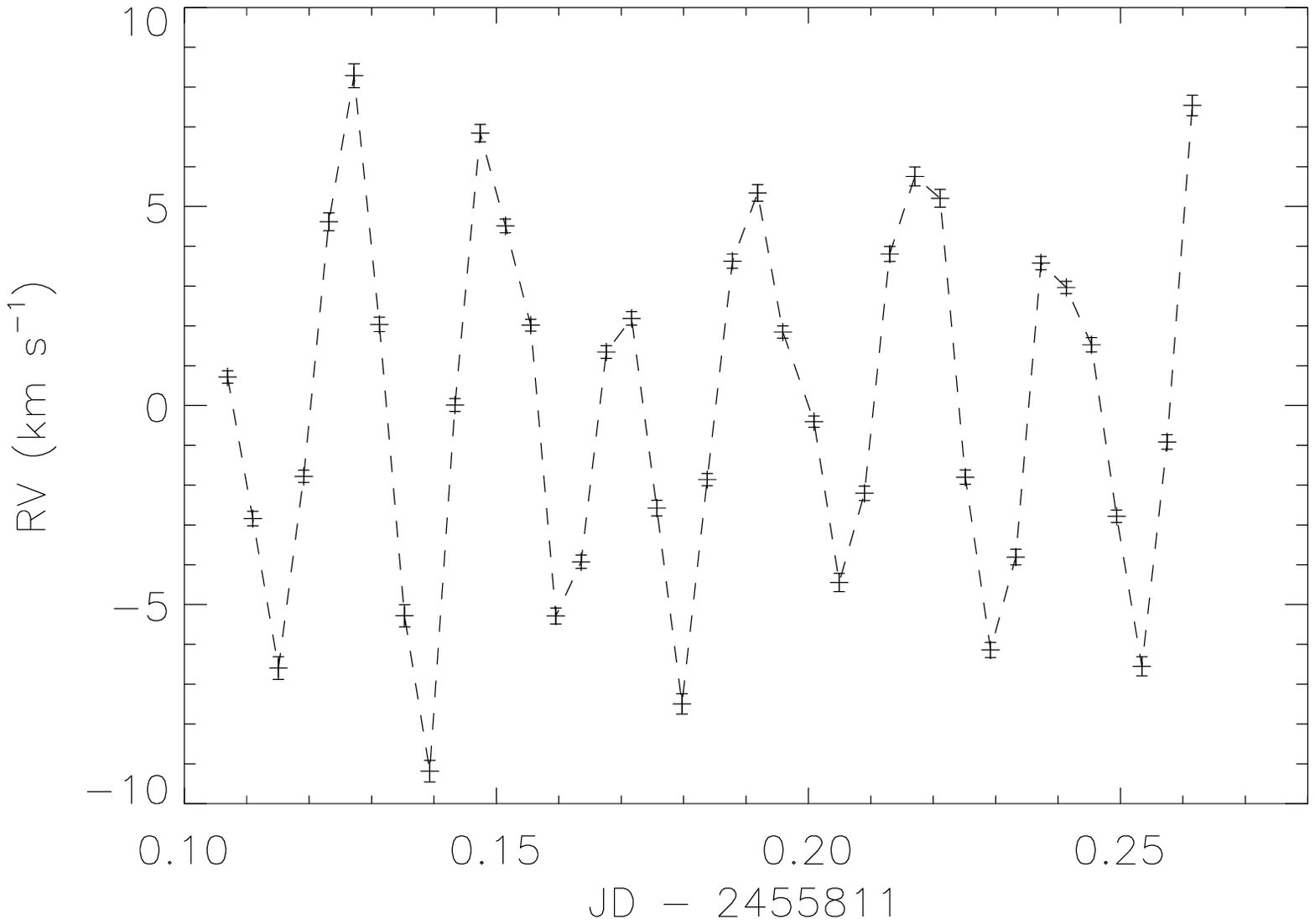}
		\includegraphics[width=0.47\textwidth,angle=0]{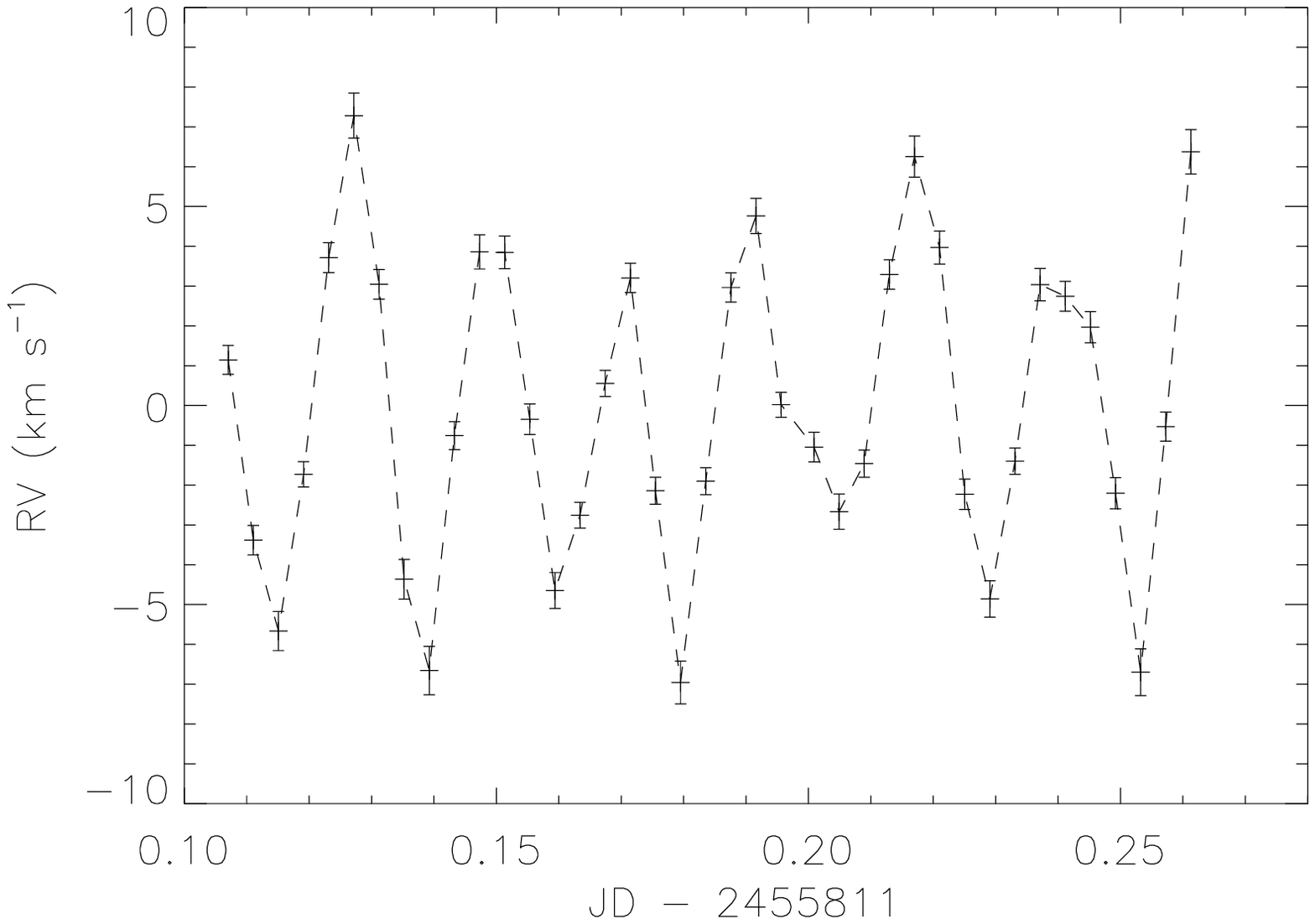}\\[1mm]
		\includegraphics[width=0.47\textwidth,angle=0]{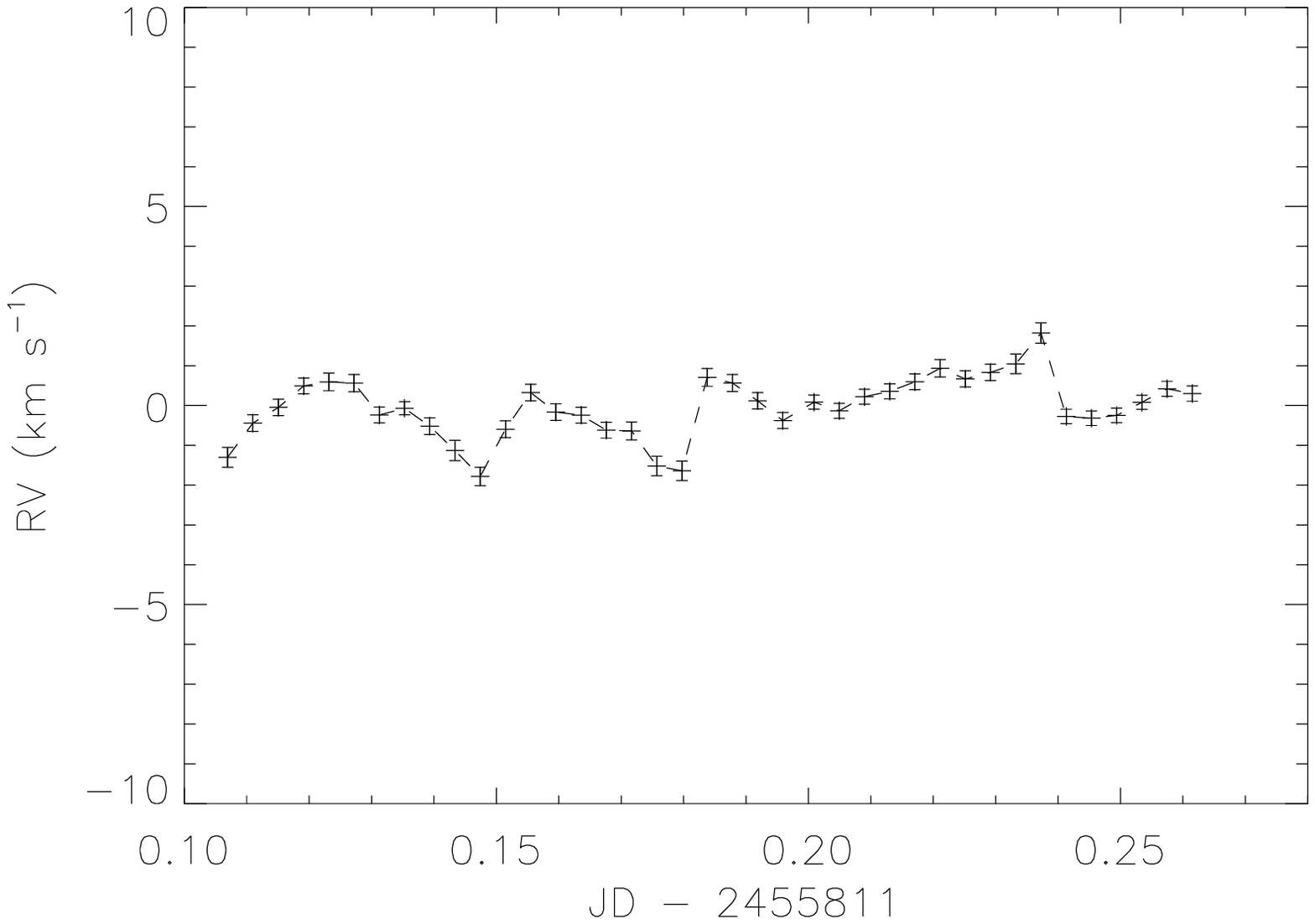}\\[1mm]
\caption{Relative radial velocities for \lsfour\ obtained with VLT/UVES from 2011 Sept 07 for 
three different wavelength regions. Top:     \ion{He}{i}\,501.6\,nm.     Middle:     `metal' absorption lines.
Bottom: 627.0--632.5\,nm telluric absorption lines. Since the sampling rate is quite low, data points are 
connected in order to guide the eye.} 
	\label{f:uves_rv}
\end{figure}

\section[]{Velocity Measurements}

Radial velocities were measured by cross-correlating each spectrum 
with a  template defined to be a mean of  all  spectra for a given observing sequence. 
For this purpose, each spectrum was normalised to an approximate continuum,  which was then subtracted.
Cross-correlation was carried out in log wavelength space. 
Velocities were measured by fitting a parabola to a region
around the maximum of the cross-correlation function (ccf) 
and customized to each wavelength region studied. The formal 
error in the position of the parabola apex was adopted as representative 
of the velocity error.

The observer's frame radial velocity of the template was obtained by cross-correlation 
with a theoretical spectrum computed for a hot subdwarf with 
atmospheric abundances of hydrogen and helium being 70\% and 30\% by number, 
solar abundances of other elements, effective temperature $\Teff=32\,000$\,K, 
surface gravity $\log g=5.5\,(\cmss)$, and microturbulent velocity $v_{\rm t}=5\kmsec$
\citep{behara06}. This is not a perfect match  to the observed spectrum, but completely satisfactory
for the purpose of velocity measurement by cross correlation.  
 Corrections to the heliocentric frame were then applied. These template
velocities ($v_{\rm T}$) are shown in Table\,\ref{t:log}. Becasue of uncertainty in the absolute wavelength
calibration, the template  velocities for the low-resolution data ($R<6\,000$) are untrustworthy
and are shown in parentheses. The {\it relative} velocities are much more reliable. 

\subsection{Radial Velocities}

 Relative velocities for the three  individual AAT observations 
are not shown since the exposure times were long
compared with the photometric variation. 

The short duration of the WHT observations and the resolution of the SAAO 
observations were inadequate to offer any prospect of detecting the pulsation at 0.022\,d, 
and show no evidence for longer-period variations (Figs.\,\ref{f:wht},\ref{f:saao}). 
Periodograms for these data showed no evidence of periodicity. 

The WiFES data are divided into two wavelength regions (418 -- 488\,nm and 490 -- 552\,nm). 
Each was studied separately.  Results are shown in Fig.\,\ref{f:anu}. In general the
two sets of velocities agree to within $2\sigma$, where $\langle\sigma\rangle\approx 2.5\kmsec$.
There is little evidence for variability in excess of $\pm 5\kmsec$ on a timescale of 0.1\,d

The UVES data were recorded in three wavelength regions (330--452, 480--575, and 583--680\,nm). 
These regions include broad Balmer, \ion{He}{i} and \ion{He}{ii} lines. They include sharp stellar 
absorption lines from both light and heavy ions. 
They also include sharp interstellar and telluric absorption lines;  
there may also be residual instrumental artefacts. The latter all produce a sharp
and stationary component in the ccf. 
The telluric absorption lines give an essentially
flat velocity curve with a maximum deviation of $\pm 2\kmsec$ 
($\langle\sigma\rangle\approx 0.25\kmsec$)   (Fig.\,\ref{f:uves_rv}). 
These velocities are consistent with small drifts in the instrumental calibration. 
The stellar absorption behaves quite differently 
and shows significant variability  with a full amplitude of up to 15 \kmsec (Fig.\,\ref{f:uves_rv}). 
Although there are  differences between velocities measured in different spectral regions and 
from different lines, the  shape of the velocity curve is maintained. Therefore {\it we conclude 
the velocity curve to be real and due to variable motion of the stellar surface in the line of sight. }

The mean heliocentric radial velocity of \lsfour\ is $-154\pm1\kmsec$. This is large, but 
not unusual for a hot subdwarf in the thick disk. We see no evidence for long-period or large-amplitude 
variability that might suggest motion within a binary system. 

\begin{figure}
	\centering
\includegraphics[width=0.47\textwidth,angle=0]{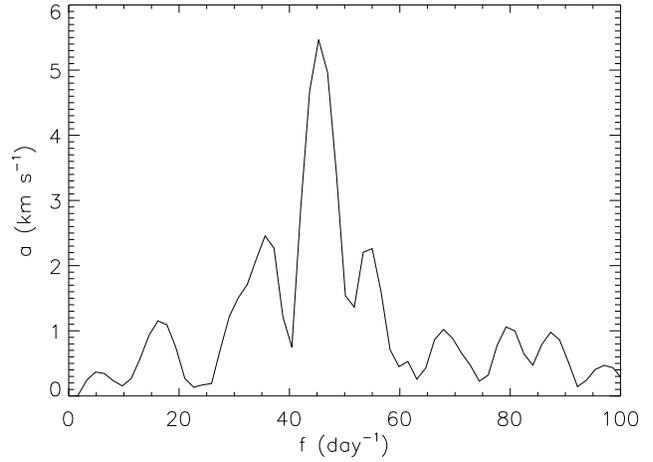}
\caption{Classical power spectrum for radial velocities obtained from the \ion{He}{i}\,501.6\,nm line 
observed with  VLT/UVES on 2011 Sept 07.} 
	\label{f:pdgram}
\end{figure}

\begin{figure}
	\centering
\includegraphics[width=0.47\textwidth,angle=0]{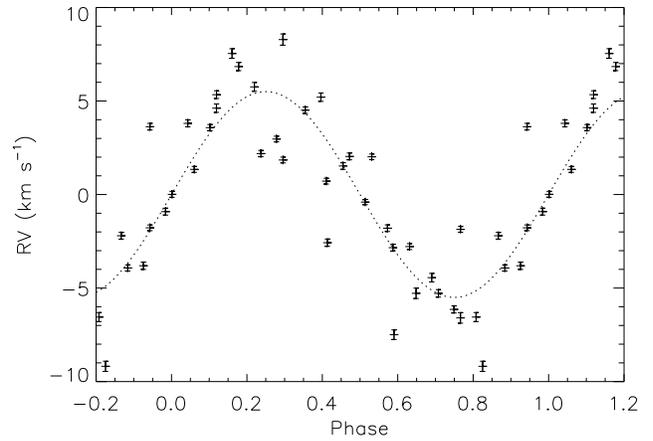}
\caption{Radial velocity curve for \lsfour\ obtained from the \ion{He}{i}\,501.6\,nm line  
phased to the principal period of 1978\,s with a semi-amplitude of 5.5\kmsec.} 
	\label{f:phase}
\end{figure}

\begin{figure}
	\centering
\includegraphics[width=0.47\textwidth,angle=0]{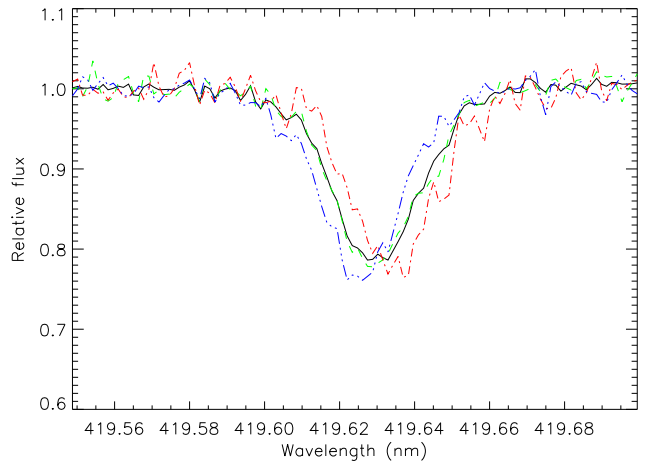}
\caption{\ion{Zr}{iv} line profiles for \lsfour\ obtained from the mean spectrum (solid -- black), 
around minimum (dot-dot-dot-dashed -- blue: $v<-5 \kmsec$), mean (dashed -- green: $|v|<1\kmsec$) and maximum (dot-dashed -- red: $v>+5 \kmsec$) 
relative radial velocity.} 
	\label{f:profs}
\end{figure}

\begin{center}
\begin{table}
\caption{Frequencies ($f$), velocity amplitudes ($a_{\rm RV}$) and residual intensity at line centre ($r_{\rm c}$)
for various wavelength windows ($\lambda$). }
\label{t:per}
\begin{tabular}{l ccc cc}
\hline
Lines  & $\lambda$  & $f$ & $a_{\rm RV}$  & $r_{\rm c}$ & $\langle\sigma_{\rm RV}\rangle$\\
 nm        &    nm         &      d$^{-1}$ &    $\kmsec$   &   &   $\kmsec$     \\ 
\hline
metals & 415.5 -- 430.0 & 43.8 &  4.6 \\
many & 490.5 -- 570.0 & 43.7 &     4.1   \\
many  & 332.0 -- 452.0 & 43.7 &    2.7 \\[1mm]
\ion{Sr}{ii}\,407.8 & 407.5 -- 407.8 & 43.8 &  3.4 & 0.90 & 0.33 \\
\ion{Sr}{ii}\,421.5 & 421.2 -- 421.6 & 43.8 &  4.3 & 0.88 & 0.32 \\
\ion{Ge}{ii}\,417.9 & 417.5 -- 417.9 & 43.8 &  5.2 & 0.86 & 0.33 \\
\ion{Ge}{ii}\,426.1 & 425.7 -- 426.0 & 43.8 &  3.7 & 0.89 & 0.32 \\
\ion{Ge}{ii}\,429.1 & 428.6 -- 429.2 & 43.8 &  1.7 & 0.95 & 0.37 \\
\ion{Y}{iii}\,404.0 & 403.6 -- 4.3.9 & 43.8 & 3.4 & 0.89 & 0.31 \\
\ion{Zr}{iv}\,413.7 & 413.4 -- 413.7 & 45.4 & 2.8 & 0.89 & 0.43 \\
\ion{Zr}{iv}\,419.8 & 419.5 -- 419.7 & 43.8  &     5.4  & 0.79 & 0.16 \\
\ion{Zr}{iv}\,431.7 & 431.3 -- 431.7 & 43.8  &     4.4  & 0.82 & 0.23 \\[1mm]
\ion{C}{ii}\,426.7 & 426.0 -- 427.0 & 43.8 &   4.2 & 0.74 & 0.25  \\
\ion{C}{iii}$\times3$  & 406.3 -- 407.0 & 43.8 &   3.9 & 0.87 & 0.25  \\
\ion{N}{ii}\,399.5 & 399.1 -- 399.5 & 43.8 &   4.5 & 0.85 & 0.28  \\[1mm]
\ion{He}{i}\,386.7 & 386.3 -- 386.8 &  43.8  & 3.7 & 0.87 & 079\\ 
\ion{He}{i}\,388.8 & 388.5 -- 389.0 &  43.8  & 1.5 & 0.60 & 0.34 \\ 
\ion{He}{i}\,396.3 & 396.0 -- 396.5 &  43.8  & 1.3 & 0.68 & 0.22 \\ 
\ion{He}{i}\,412.1 & 411.6 -- 412.2 &  43.8  & 5.1 & 0.75 & 0.77 \\ 
\ion{He}{i}\,438.8 & 438.0 -- 439.3  & 43.8  & 5.2 & 0.74  & 1.63 \\
\ion{He}{i}\,447.1 & 445.0 -- 449.0  & 43.8  & 3.8 & 0.55  & 1.41 \\
\ion{He}{i}\,501.6 & 500.0 -- 502.5 &  43.8  & 5.5 & 0.66 & 0.20 \\ 
\ion{He}{i}\,667.8 & 666.0 -- 669.0 &  43.8  & 5.0 & 0.59 & 0.47  \\[1mm]
H$\alpha$ &   655.5 -- 656.5   & 45.3   & 1.4   & 0.69 & 1.63  \\
H$\beta$  &   484.0 -- 488.0   & 43.7   &  4.9  & 0.61 & 1.12  \\
H$\gamma$ &  432.0 -- 435.5   & 43.7  & 2.7  & 0.60 & 1.77  \\[1mm]
telluric &  627.0 -- 632.5 &  34  &  0.3 & 0.75  \\
\hline
$\langle \sigma \rangle$ &        &   6.5   &  0.1 &  0.01 \\ 
\hline
\end{tabular}
\end{table}
\end{center}

\subsection{Period}

For each set of velocities measured from the UVES data, we computed the classical Fourier 
power spectrum and measured the frequency and semi-amplitude (as square root of power) 
of the highest peak  (Fig.\,\ref{f:pdgram}, Table\,\ref{t:per}). Owing to the short duration 
 of the data series, ($T=0.155$\,d), the
frequency resolution is low ($\Delta f = 1/T = 6.5\,{\rm d^{-1}}$). However, the
frequency of the highest peak is clearly consistent with that of the
largest-amplitude oscillation in the photometry of both \citet{green11} and \citet{jeffery11.ibvs}. 
All lines for which a good solution was obtained yielded the same frequency (43.8\,d$^{-1}$).
The phases were the same to within $\pm0.02$ cycles. 

Figure\,\ref{f:phase} shows the radial velocities for one line (\ion{He}{i}\,501.6\,nm) 
phased on the dominant period (0.023\,d).  There remains substantial scatter about the best-fit sine curve,  in excess of the 
formal errors on the velocities. Given the agreement between velocities obtained 
at different wavelengths, and the stability of the telluric velocity data, 
there is no reason to consider  the errors to be seriously underestimated. 
We posit that {\it there is unresolved motion due to oscillations at other
periods} as indicated by photometry.

\begin{figure}
	\centering
\includegraphics[width=0.47\textwidth,angle=0]{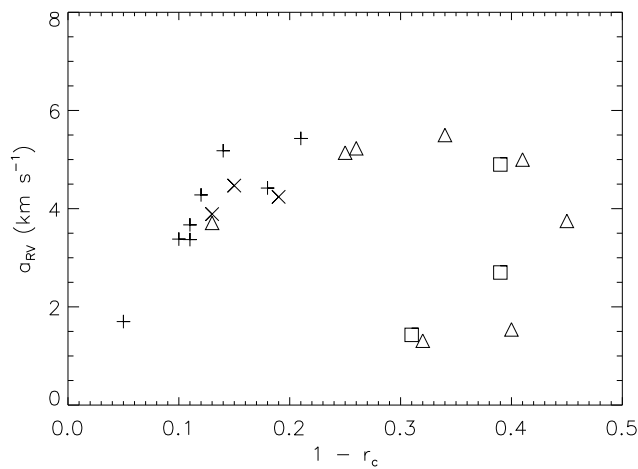}
\caption{Radial velocity amplitude as a function of line depth. Symbols represent different elements: 
Sr, Ge, Y and Zr (+), C and N ($\times$), He ($\triangle$) and H ($\square$). 
} 
	\label{f:amps}
\end{figure}

\subsection{Amplitude}

It became clear that the amplitude of the velocity variation $a_{\rm RV}$ is sensitive to the wavelength
window selected for cross-correlation (Table\,\ref{t:per}). 
The presence of interstellar lines or instrumental
artefacts was indicated by a small sharp peak in the ccf, but by fitting the ccf peak 
over a large enough window, this was not a significant factor.  Large spectral windows 
including many broad lines ({\it e.g.}  332--452\,nm) would give small amplitudes, more 
restricted windows  containing only sharp metal lines would give larger amplitudes 
({\it e.g.}  415--430\,nm), whilst some individual lines would yield still larger amplitudes
({\it e.g.} \ion{He}{i}\,501.6\,nm, \ion{Zr}{iv}\,419.8\,nm). 

For these UVES data, the  cross-correlation approach yields  high quality radial velocity  information 
from  individual lines in which the central depth $r_{\rm c}$ is at least  5\% below continuum. 
Results for selected lines are shown in  Table\,\ref{t:per}, which also includes the
mean velocity error ($\langle \sigma_{\rm RV} \rangle$) for each line measured.  There is a general trend that, 
for sharp lines, the velocity  amplitude increases with line depth (Fig.\,\ref{f:amps}). 

For broad lines, especially the Balmer series and the diffuse \ion{He}{i} lines, the amplitudes
were frequently much lower ($\approx 1 - 2 \kmsec$), principally  because the ccf peak  
is also very broad and therefore poorly defined.   However, the same period of 0.023\,d 
was generally recovered, except in the case of the Balmer lines. 

One interpretation of the increase in velocity amplitude with central depth is that 
the pulsation amplitude is a function of position in the stellar atmosphere. Strong lines are
formed at lower optical depths and, hence, higher in the atmosphere where densities are lower.
 An outward  running wave will increase in amplitude as it propagates into less dense material, 
exactly as is observed in the sharp metal lines. Similar phenomena are seen, for example, 
in rapidly-oscillating Ap stars \citep{kurtz07}.  In the case of \lsfour\ the situation is complicated by the apparent
super-abundances of zirconium, strontium, etc., since it has been argued that the atmosphere
must be chemically stratified. More detailed analysis of these and similar data should enable us 
to establish the relative depth of the chemically stratified layers relative to the  velocity 
gradient in the photosphere.

\subsection{Line Profiles}

For large-amplitude radially pulsating stars,  velocity shifts are usually associated with line-profile
shifts; owing to the centre-to-limb contrast in radial velocity,  there is usually some 
absorption at zero-velocity, whilst the line centre shifts back and forth producing asymmetric
profiles at minimum and maximum radial velocity \citep{montanes01,jeffery13.fuji2}. Non-radial
pulsations are often associated with line-profile variations, but usually in rapidly-rotating 
stars where temperature variations across the surface produce more or less flux at different 
velocities. We checked for line profile changes by coadding spectra around minimum
velocity ($\delta v<-5\kmsec$), mean velocity ($-1<\delta v<+1 \kmsec$) and maximum velocity
($\delta v>5\kmsec$). No asymmetries were identified ({\it cf.} Fig.\,\ref{f:profs}).

The use of a ccf template constructed as a simple mean allows for the possibility of some
velocity smearing. In addition, the AAT/UCLES spectra used previously for atmospheric 
analysis were obtained with long exposures ({\it cf.} \citet{naslim11}). 
The question arises whether either the 
template could be sharpened, or whether previous analyses overlooked pulsation
-broadening of the line profiles that led to an overestimate of the rotational broadening. 
We found no evidence that the spectrum of \lsfour\ could be further sharpened
(Fig.\,\ref{f:profs}), or that the \citep{naslim11} measurement of $v \sin i < 2\kmsec $ 
should be revised  downward.   

\subsection{Light -- velocity amplitude ratio}

\citet{green11,jeffery11.ibvs} give the amplitude of the light variation in the optical as 0.27\% or 
$\delta V \approx3$ mmag. 
Taking the maximum amplitude of the radial velocity amplitude to be 
$\delta v\approx6\,\kmsec$,  the ratio of the 
observed light-amplitude to velocity amplitude in mmag\,km$^{-1}$\,s is thus 
$\delta V/\delta v\approx 0.5$. This ratio assumes that the amplitude of the
1953\,s oscillation does not vary significantly over time. 

We can simulate the observable light and amplitude ratios for non-radial modes of
different radial degree $\ell$ and azimuthal number $m$ {\bf using the  
surface codes} {\sc bruce} and {\sc kylie} \citep{townsend03, ramachandran04} and a grid
of theoretical spectra for a composition appropriate to \lsfour\ \citep{behara06,naslim11}. 
Assuming effective temperature  $\Teff=36\,000$\,K,  surface gravity $\log g=5.6 (\cmss)$ 
and equatorial rotation velocity $v_{\rm eq}=2\kmsec$  \citep{naslim11}, 
and {\bf also assuming the adiabatic approximation, } a polar radius 
of 0.2\Rsolar\  (implying a mass $M\approx0.5\Msolar$),  inclination $i=80^{\circ}$, 
pulsation period 1950\,s and an arbitrary surface velocity amplitude, 
we can compute a time series of
theoretical spectra for a given mode $\ell, m$.  These can be analysed in exactly the same way
as the UVES spectra to give both the apparent velocity amplitude and, also, the apparent flux
amplitude in a given wavelength region. The ratio of surface velocity  to apparent velocity 
amplitude is a function of $\ell, m$ and $i$, so we cannot infer anything directly 
from $\delta v$. However, the ratio $\delta V/\delta v$ is relatively invariant to $i$
\citep{stamford81}  and also to the surface velocity amplitude.

 We computed theoretical  $\delta V/\delta v$ for $\ell=0,1,2$ and $m=0,\ldots,\ell$ for the dominant
mode in \lsfour, obtaining $\delta V/\delta v\approx 2$\,mmag\,km$^{-1}$\,s for the
radial mode $\ell=0$ and  $\delta V/\delta v\approx 0.5 - 0.6$\,mmag\,km$^{-1}$\,s for the
non-radial $\ell=1$ and $\ell=2$ modes. This {\it strongly} argues {\it against} the 
1953\,s period being a radial mode, in complete agreement with the argument from pulsation
theory given by \citet{green11}.  

\section{Conclusions}

We have conducted a search for radial velocity variability in the 
helium-rich hot subdwarf  \lsfour: the pulsating zirconium star.

We have demonstrated that at least one periodic variation identified photometrically in \lsfour\ is
also observable as a radial velocity variation with a semi-amplitude in excess of 5\kmsec. 
This provides strong evidence that both  are due to an oscillatory 
motion of the surface and are hence most likely due to  pulsation. 
That is a purely empirical conclusion. Additional evidence from the light/velocity amplitude 
ratio  argues that  the pulsation cannot be a radial mode, whilst 
supporting  arguments from pulsation theory \citep{green11} confirm that 
the observed pulsations are due  to a non-radial gravity mode.

We have also demonstrated that differential motion within the photosphere can be resolved
by studying lines of different depths, as these probe different levels of the photosphere.

To learn more about \lsfour\ from these oscillations, the radial velocity measurements
must be repeated over a  longer period of time, and should be supported be
multi-wavelength photometry. Higher frequency resolution would allow the 
relative phases of different lines to be measured and show how the oscillation propagates
through the photosphere.  Ultraviolet spectrophotometry would 
 help to identify the modes of oscillation, as well as to 
address the effective temperature question. Theoretical models need to be developed
to  interpret the spectrum, incorporating the non-radial oscillations, chemical stratification, 
and including differential vertical motion. In particular, the profiles and behaviour of the 
\ion{He}{i} and hydrogen Balmer lines deserve further study. 

\section*{Acknowledgments}

The Armagh Observatory is funded by direct grant-in-aid from the Northern Ireland Dept of Culture, Arts and Leisure.
Observing travel for AA was funded by a PPARC grant. 
CSJ is indebted to Suzanna Randall for encouraging him to complete this study. 

\bibliographystyle{mn2e}
\bibliography{ehe}

\label{lastpage}

\end{document}